\documentclass[
  reprint,
  amsmath,amssymb, aps,
  prb,
]{revtex4-1}

\usepackage{graphicx}
\usepackage{dcolumn}
\usepackage{hyperref}
\usepackage{cancel} % goes to zero arrows
\usepackage[normalem]{ulem}
\usepackage[usenames, dvipsnames]{color}
\usepackage[normalem]{ulem}
\begin{document}
\newcommand{\kb}{$\mathbf{k}$}

\preprint{APS/123-QED}

\title{Efficiency of Generalized Regular \kb-point Grids}
% Force line breaks with\\
%\thanks{}

\author{Wiley S. Morgan, Jeremy J. Jorgensen, Bret C. Hess, Gus
  L. W. Hart} \affiliation{Department of Physics and Astronomy,
  Brigham Young University, Provo, Utah, 84602, USA}

% \collaboration{\noaffiliation}

\date{\today}
\begin{abstract}
Most DFT practitioners use regular grids (Monkhorst-Pack, MP) for
integrations in the Brillioun zone. Although regular grids are the
natural choice and easy to generate, more general grids whose
generating vectors are not merely integer divisions of the reciprocal
lattice vectors, are usually more efficient.\cite{wisesa2016efficient}
We demonstrate the efficiency of \emph{generalized regular} (GR) grids
compared to Monkhorst-Pack (MP) and \emph{simultaneously commensurate}
(SC) grids. In the case of metals, for total energy accuracies of one
meV/atom, GR grids are 60\% faster on average than MP grids and 20\%
faster than SC grids. GR grids also have greater freedom in choosing
the \kb-point density, enabling the practitioner to achieve a target
accuracy with the minimum computational cost.
\end{abstract}

%\pacs{Valid PACS appear here}% PACS, the Physics and Astronomy %
% \keywords{Suggested keywords}%Use showkeys class option if keyword
\maketitle

\clearpage
\newpage
\clearpage
\onecolumngrid

\clearpage
\begin{figure}[t]
  \centering
  \includegraphics[width=18cm]{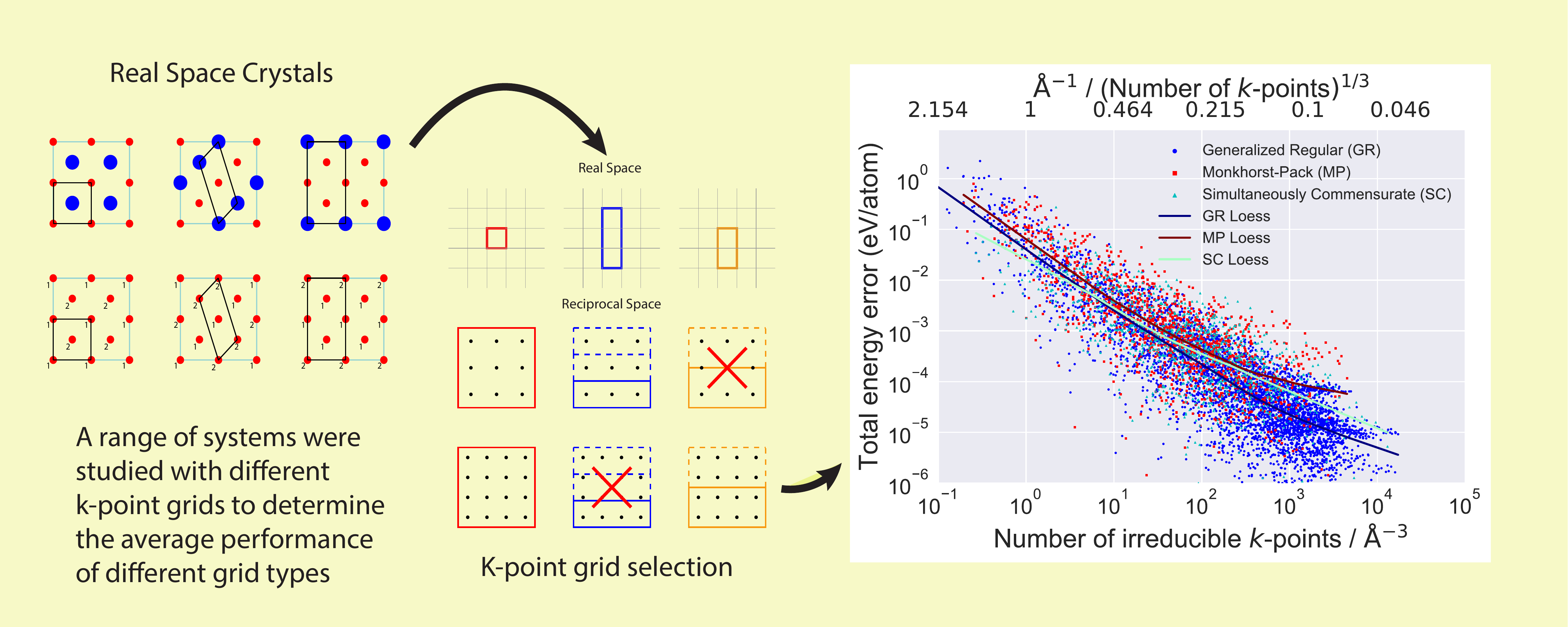}
  \label{fig:graphical_abstract}
\end{figure}

\newpage
\clearpage

\section{Highlights}
\begin{itemize}
  \item{The efficiency of Generalized Regular, Simultaneously Commensurate, and
    Monkhorst-Pack \kb-point grids are compared.}
  \item{Generalized Regular \kb-point grids are found to be 20\% more efficient
    than Simultaneously Commensurate grids.}
  \item{Generalized Regular \kb-point grids are found to be 60\% more efficient
    than Monkhorst-Pack grids.}

\clearpage    
\end{itemize}
\twocolumngrid

\section{Introduction}
High throughput materials design has become an effective route to
material discovery with many successes already documented
\cite{greeley2006computational, gautier2015prediction,
  oliynyk2017discovery, chen2012carbonophosphates,
  hautier2011phosphates, jahne2013new, moot2016material,
  aydemir2016ycute, zhu2015computational, chen2016understanding,
  ceder1998identification, yan2015design, bende2017chemical,
  mannodi2017scoping, sanvito2017accelerated,
  yaghoobnejad2016combined, hautier2013identification, bhatia2015high,
  ceder1998identification, johannesson2002combined,
  stucke2003predictions, curtarolo2005accuracy, matar2009first,
  ceder2011recharging, sokolov2011computational, ulissi2017machine,
  levy2009new, ma2013improved, yang2012search, chen2012synthesis,
  kirklin2013high}. The creation of large material databases is the
first step in high throughput approaches \cite{curtarolo2012aflow,
  saal2013materials, jain2013commentary, NOMAD,
  landis2012computational, hachmann2011harvard, hummelshoj2012catapp,
  de2015database, de2015charting, cheng2015accelerating,
  gomez2016design, chan2015combinatorial, tada2014high,
  pilania2013accelerating, yan2015material, ramakrishnan2014quantum,
  hachmann2014lead, lin2012silico, armiento2014high,
  senkov2015accelerated}. Computationally expensive electronic
structure calculations generate the data for the databases and limit
the extent to which data analysis tools, such as machine learning, can
be applied. Increasing the speed of these calculations has the
potential to significantly increase the size of these databases and
the impact of material predictions.

Most electronic structure codes perform numerical integrals over the
first Brillouin zone, which converge extremely slowly in the case of
metals. Dense sampling of the Brillouin zone, required for high
accuracy, is computationally expensive, especially when implementing
hybrid functionals or perturbative expansions in density functional
theory (DFT) \cite{berland2017enabling}. High accuracy is important
because the energies of competing phases are often similar and even
small errors can affect the prediction of stable materials.

Methods for \kb-point selection have not changed much since Monkhorst
and Pack published their influential paper over 40 years ago
\cite{monkhorst1976special}. Their method was quickly accepted by the
community due to its simplicity and ability to generalize previous
methods \cite{baldereschi1973mean, chadi1973special}. Sampling methods
that improve upon Monkhorst-Pack (MP) grids have been far less
prevalent \cite{froyen1989brillouin, moreno1992optimal,
  wisesa2016efficient, CanES}.

In this paper, we compare the \kb-point selection method promoted by
Wisesa, McGill, and Mueller\cite{wisesa2016efficient} (WMM) to the
standard MP grids and to another common method in the alloy community,
which we refer to as \textit{simultaneously commensurate} (SC)
grids. This paper serves to reinforce and quantify the claims made by
WMM, as applied to calculations typically used for alloys and for some
high-throughput studies.

\section{Background}
Over the past 40 years, only a few \kb-point selection methods have
been proposed in the literature \cite{baldereschi1973mean,
  chadi1973special, monkhorst1976special, froyen1989brillouin,
  moreno1992optimal, wisesa2016efficient}. Many of these so-called
special point methods have focused on selecting points that accurately
determined the mean value of a periodic function defined over the
Brillouin zone because the integral of a periodic function over one
period is simply its mean value. Other factors that have been
considered in developing special point methods are selection of grids
with a consistent density in each direction and full exploitation of
symmetry.

Baldereschi introduced the \textit{mean-value point} of the Brillouin
zone \cite{baldereschi1973mean}, the first special point method. In
this approach, the periodic function to be integrated is written as a
Fourier expansion:
\begin{equation}
f(\mathbf{k}) = \sum_{n=0}^{\infty} c_n e^{i \mathbf{k} \cdot \mathbf{R}_n},
\end{equation}
where $\mathbf{k}$ is the wavevector, $c_n$ is the $n$-th expansion
coefficient, and the sum is over over all lattice points
$\mathbf{R}_n$. Baldereschi noted that the integral of $f(\mathbf{k})$
within the first Brillouin zone (i.e., over one period of $f($\kb$)$),
is proportional to the leading coefficient, $c_0$, in the Fourier
expansion,
\begin{equation}
\int_\text{BZ} f(\mathbf{k})\, d\mathbf{k} = \frac{(2\pi)^3}{\Omega} c_0,
\end{equation}
where $\Omega$ is the volume of the reciprocal cell. He replaced the
analytic integral of the periodic function with a numeric integral
(sum over $j$ in Eq.~\ref{eq:special_points})---equivalent in the
limit of infinite sampling points---and replaced the periodic function
with its infinite Fourier expansion (sum over $n$ in
Eq.~\ref{eq:special_points}):
\begin{align} \label{eq:special_points}
  \int_\text{BZ} f(\mathbf{k}) &= \sum_{j=0}^{\infty} w_j
  f(\mathbf{k}_j) \nonumber \\
  &=
  \sum_{j=0}^{\infty} w_j \sum_{n=0}^{\infty} c_n e^{i \mathbf{k}_j
    \cdot \mathbf{R}_n} \nonumber \\
  &=
  \sum_{j=0}^{\infty} w_j (c_0 + c_1 e^{i \mathbf{k}_j \cdot
    \mathbf{R}_1} + \dots )\nonumber \\%% + c_2
  %% e^{i \mathbf{k}_j \cdot \mathbf{R}_2} + \dots) \nonumber \\
  &= \sum_{j=0}^{\infty} w_j c_0 + \sum_{j=0}^{\infty} w_j c_1 e^{i \mathbf{k}_j \cdot
    \mathbf{R}_1} + \dots, %% \sum_{j=0}^{\infty} w_j c_2 e^{i \mathbf{k}_j \cdot \mathbf{R}_2} + \dots.
\end{align}
where $w_j$ is the integration weight of the $j$-th \kb-point. In the
final step of Eq.~\ref{eq:special_points}, each term (sum over $j$) is
a numeric integral of the $n$-th basis function in the Fourier
expansion of $f(\mathbf{k})$ (denoted as $I_n$ in what
follows). Baldereschi's method selected \kb-points so that the leading
terms after $c_0$ integrate to zero:
\begin{align*}
  \int_\text{BZ} f(\mathbf{k}) &= \sum_{j=0}^{\infty} w_j c_0 +
  \sum_{j=0}^{\infty} w_j c_1 e^{i \mathbf{k}_j \cdot \mathbf{R}_1} + \\
  & \hspace{.3in} \sum_{j=0}^{\infty} w_j c_2 e^{i \mathbf{k}_j \cdot \mathbf{R}_2} +
  \dots \\
  &= I_0 + \cancelto{0}{I_1} + \cancelto{0}{I_2} + \mathcal{O}(I_3),  
  \\ &\approx c_0 \, \sum_j w_j.
\end{align*}

\begin{figure}[t]
  \centering
  \includegraphics[width=8cm]{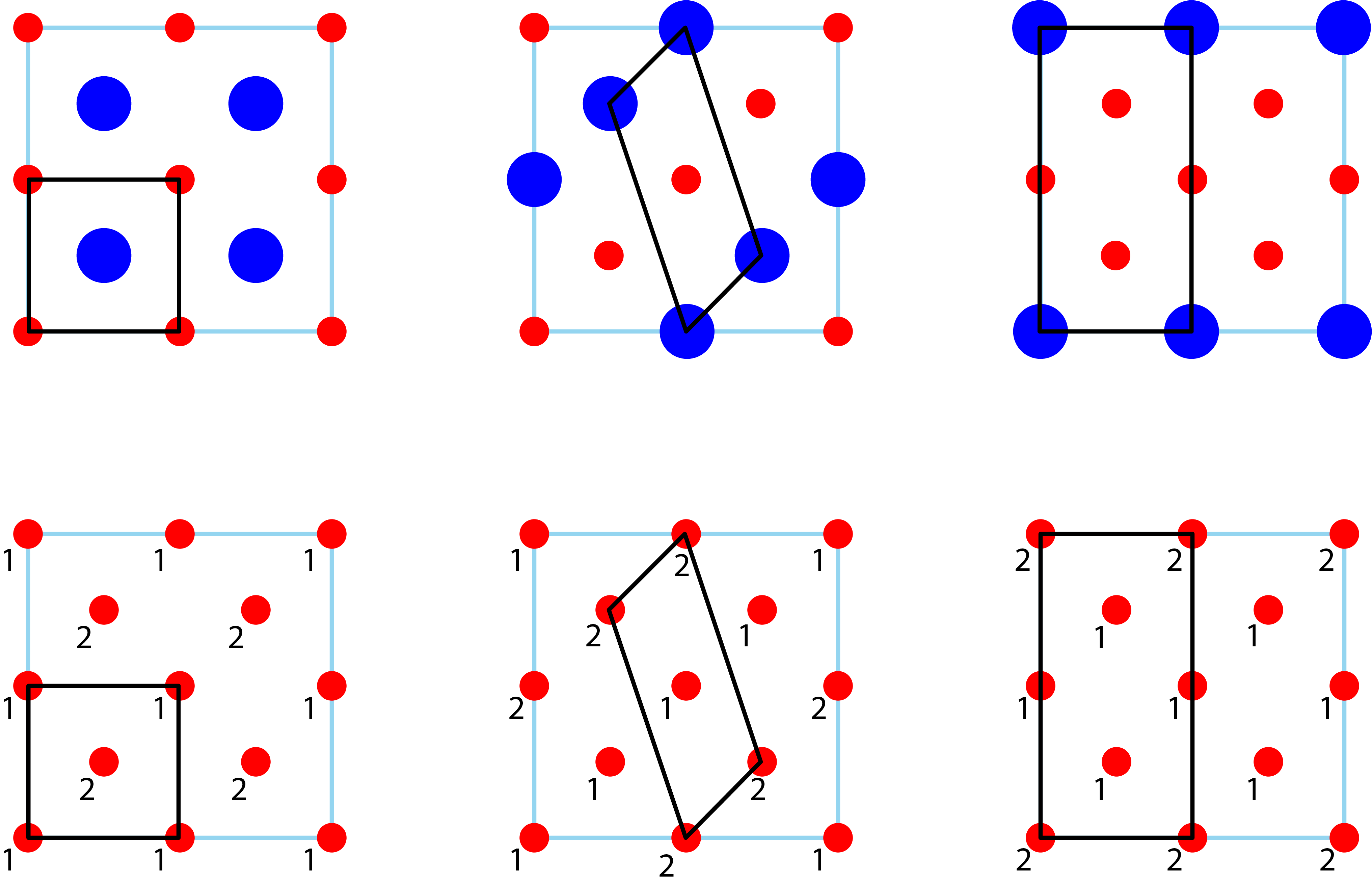}
  \caption{In order to isolate the effect of the Brillouin zone shape
    and size on total energy error when comparing crystal structures
    of different shapes and sizes (top row), the energy of supercells
    (bottom row) crystallographically equivalent to single element,
    primitive cells were compared. The total energy per atom should be
    the same for all equivalent cells.}
  \label{fig:supercells_same_atoms}
\end{figure}

This is an accurate approximation when the Fourier coefficients
converge rapidly to zero, as is the case with insulators and
semiconductors. Baldereschi's approximation is ineffective for metals
because the integral over the occupied parts of the band structure has
discontinuities, and the Fourier series converges very slowly.

Chadi and Cohen extended the mean-value point by introducing
\textit{sets} of \kb-points whose weighted sum eliminated the
contribution of a greater number of leading basis
functions\cite{chadi1973special}. Their sets of \kb-points could be
made as dense as desired.

The most popular \kb-point selection method was created by Monkhorst
and Pack \cite{monkhorst1976special} (MP). They established a grid of
points that generalized both the mean-value point of Baldereschi and
its extension by Chadi and Cohen and which was equivalent to points
used by Janak et al.\cite{janak1971gilat} MP grids are given by the
relation
\begin{equation}
\mathbf{k}_{p r s} = u_p \mathbf{b}_1 + u_r \mathbf{b}_2 + u_s
\mathbf{b}_3
\end{equation}
where $\mathbf{b}_1$, $\mathbf{b}_2$, and $\mathbf{b}_3$ are the
reciprocal lattice vectors, $u_p = (2p - q -1)/2q$ for
$p = 1, 2, \ldots, q$, and $q$ an integer that determines the grid
density. The same relation holds for $u_r$ and $u_s$. In other words,
the generating vectors of MP grids are simply integer divisions of the
reciprocal lattice vectors.

Froyen generalized the MP points, which he called \textit{Fourier
  quadrature points}, by eliminating the restriction that the vectors
that defined the grid be parallel to the reciprocal lattice vectors
\cite{froyen1989brillouin}. However, he did require the grid to be
commensurate with the reciprocal lattice and to have the full
point-group symmetry of the crystal.

\begin{figure}[t]
  \centering
  \includegraphics[width=8cm]{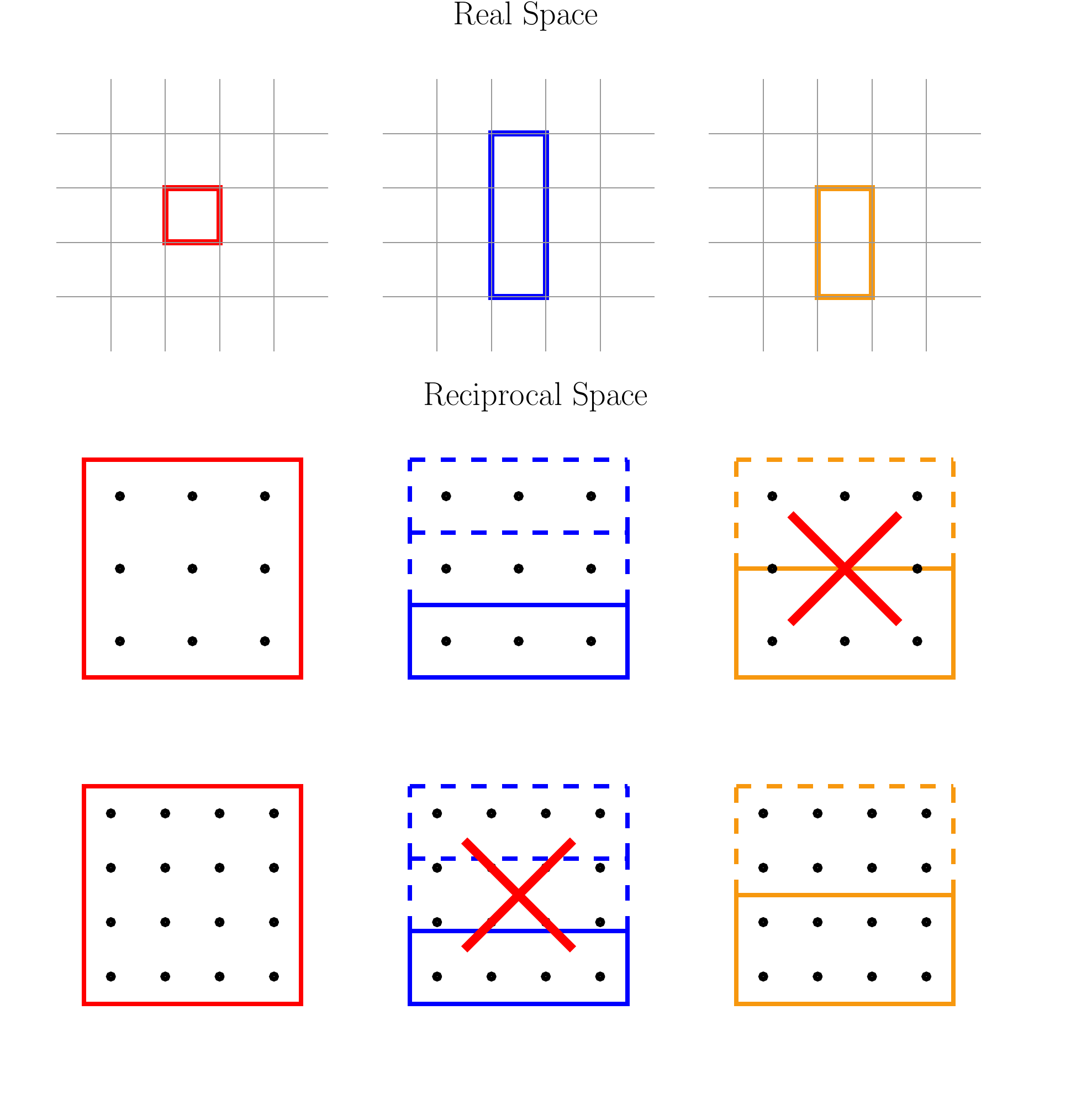}
  \caption{An example of simultaneously commensurate grids. The cells
    for each crystal are shown in both real and reciprocal space. In
    reciprocal space, we include two \kb-point grids of different
    density. Simultaneously commensurate grids eliminate systematic
    \kb-point error (between two commensurate structures) by using the
    same grid for both the parent cell (red cell) and the supercells
    (yellow and blue). However, some grids may not be
    allowed (crossed out) for a given supercell because they are
    incommensurate with the reciprocal cell.
  \label{fig:simultaneously-commensurate}} 
\end{figure}

Moreno and Soler\cite{moreno1992optimal} introduced the idea of
searching for \kb-point grids with the fewest points for a given
length cutoff---a parameter that characterized the quality of the grid
and was closely related to the \kb-point density. Their method
constructs superlattices of the real-space primitive lattice. The
\emph{dual} of the superlattice vectors form the \kb-point grid
generating vectors. By selecting superlattices that maximize the
minimum distance between lattice points (i.e., by choosing fcc-like
superlattices), they obtain \kb-point grids that are bcc-like. Grids
that are bcc-like have the smallest integration errors at a given
\kb-point density. (This is evident in Fig.~\ref{fig:si_convergence}.)
Moreno and Soler further improved Brillouin zone sampling by finding
the offset of the origin that maximized the symmetry reduction of the
grid.

In their recent paper, WMM point out that the lack of popularity of
Moreno and Soler's approach is due to the computational expense of
calculating many Froyen grids and searching for the ones with the
highest symmetry reduction. They used the term \emph{Generalized
  Monkhorst-Pack} (GMP) grids to refer to Froyen grids with the
highest symmetry reduction for a given \kb-point density.  We refer to
these grids as \emph{Generalized Regular} (GR) grids since they are
simply generalizations of the regular grids used in finite element,
finite difference, and related methods. WMM precalculated the grids,
and stored the ones with the highest symmetry reduction in a database
that can be accessed via an internet request.

 \begin{figure}[t]
  \centering
  \includegraphics[width=8cm]{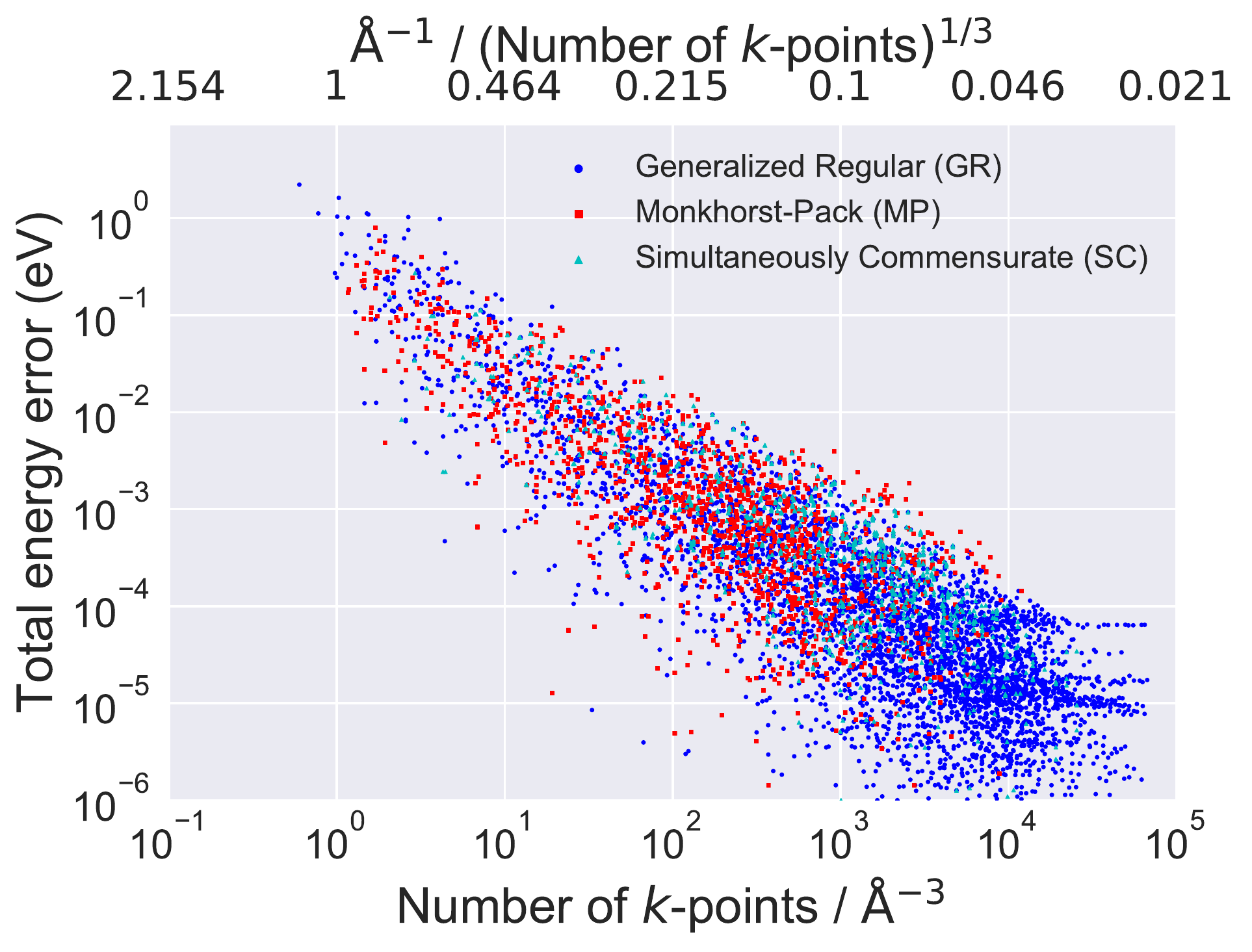}
  \caption{Total energy convergence by grid type. Note that
    the size of the convergence envelope gets bigger with higher
    \kb-point densities.}
  \label{fig:mueller_comb}
\end{figure}

\section{Methods}

We compare the total energy errors of MP, SC, and GR grids for
different \kb-point densities over calculations of nine different
elements (all of which are metallic), many cell shapes, and cell sizes
from 1--14 atoms. In total we compare errors across more than 7000
total energy calculations. One \kb-point grid is considered superior
to another if it requires a smaller irreducible \kb-point density to
reach a specific accuracy target (for example $10^{-3}$ eV/atom). The
method that requires the smallest irreducible \kb-point density is the
one we regard as best suited for high throughput and machine learning
applications.

To isolate error arising from \kb-point integration, the different
cells were crystallographically equivalent to single element,
primitive cells, as illustrated in
Fig. \ref{fig:supercells_same_atoms}. We did this to study how
\kb-point error depends on the Brillouin zone shape and size; this is
an important consideration in high-throughput studies where total
energy differences between competing phases are critical.

The grid types we compared were MP grids (generated by AFLOW's
algorithm \cite{curtarolo2012aflow}), SC
grids\cite{froyen1989brillouin} (examples of SC grids can be found in
Fig.~\ref{fig:simultaneously-commensurate}, details of SC grid
  generation can be found in the appendix), and GR grids (generated by
querying WMM's \kb-point
server).\footnote{\href{http://muellergroup.jhu.edu/K-Points.html}{http://muellergroup.jhu.edu/K-Points.html}}
We ran DFT calculations using the Vienna Ab-initio Simulation Package
(VASP) \cite{kresse1993ab, kresse1996efficiency, kresse1994ab,
  kresse1996efficient} on nine monoatomic systems---Al, Pd, Cu, W, V,
K, Ti, Y, and Re---using PAW PBE
pseudopotentials.\cite{blochl1994projector, kresse1999ultrasoft} The
supercells of cubic systems varied between 1--11 atoms per cell, while
the hexagonal close packed (HCP) systems had 2--14 atoms per cell. We
used VASP 4.6 for all calculations.\footnote{For SC grids an
  independent $ \mathbf {k}$-point folding algorithm was used due to
  an occasional bug in VASP 4.6's folding algorithm. This bug has been
  fixed in version 6 of VASP.}  For MP and SC grids the target number
of \kb-points extended from 10--10,000 unreduced \kb-points The range
of \kb-points for GR grids was 4--150,000
\kb-points.\footnote{Calculations with MP grids with more than about
  10,000 unreduced points were not used. These calculations were
  problematic due to a number of problems, including memory
  constraints and errors during the \kb-point folding.}

The converged total energy, the energy taken as the error-free
``solution'' in our energy convergence comparisons, was the
calculation with the highest \kb-density for each system. Because MP
and SC grids are difficult to generate at comparable densities, GR
grids were used to generate the converged total energy.

\begin{figure}[t]
  \centering
  \includegraphics[width=8cm]{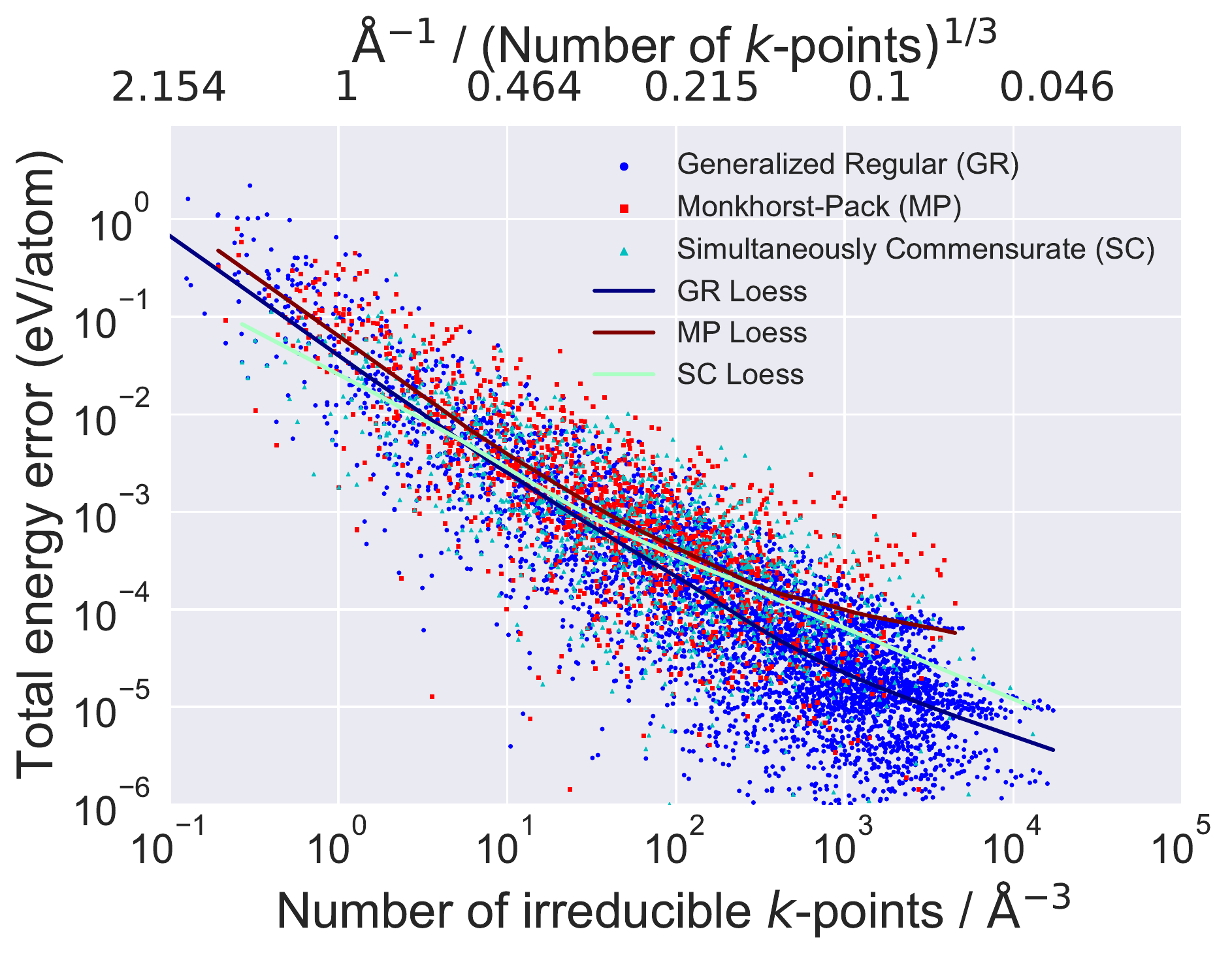}
  \caption{Total energy convergence with respect to the irreducible
    \kb-point density.  By looking at the irreducible \kb-point
    density the efficiency of the different
    grids can be distinguished. Loess smoothing was also employed to
    determine the average efficiency of the
    grids.}
  \label{fig:all_metal_comb}
\end{figure}

\section{Results}

In Fig. \ref{fig:mueller_comb}, we show the convergence for the MP,
SC, and GR grids with respect to the \kb-point density, i.e.,
\kb-points/\AA$^{-3}$. The first thing to note is the large spread in
the convergence. This spread reduces the reliability of
high-throughput databases and is perhaps higher than one might
expect. Note that the size of the total energy convergence envelope
(variance) gets bigger with increasing \kb-point
densities. Additionally it can be seen that each method has the same
variance at all \kb-point densities.

In order to quantify the efficiency of GR grids
relative to SC and MP grids, we studied the rate of energy convergence
with respect to the irreducible \kb-point density, i.e., the number of
\emph{irreducible} \kb-points divided by the volume of the reciprocal
cell in \AA$^{-3}$ (shown in Fig.~\ref{fig:all_metal_comb}). Given the
amount of scatter in the plot, we performed loess regression to create
trend lines for each grid type.

\begin{figure}[t]
  \centering
  \includegraphics[width=8cm]{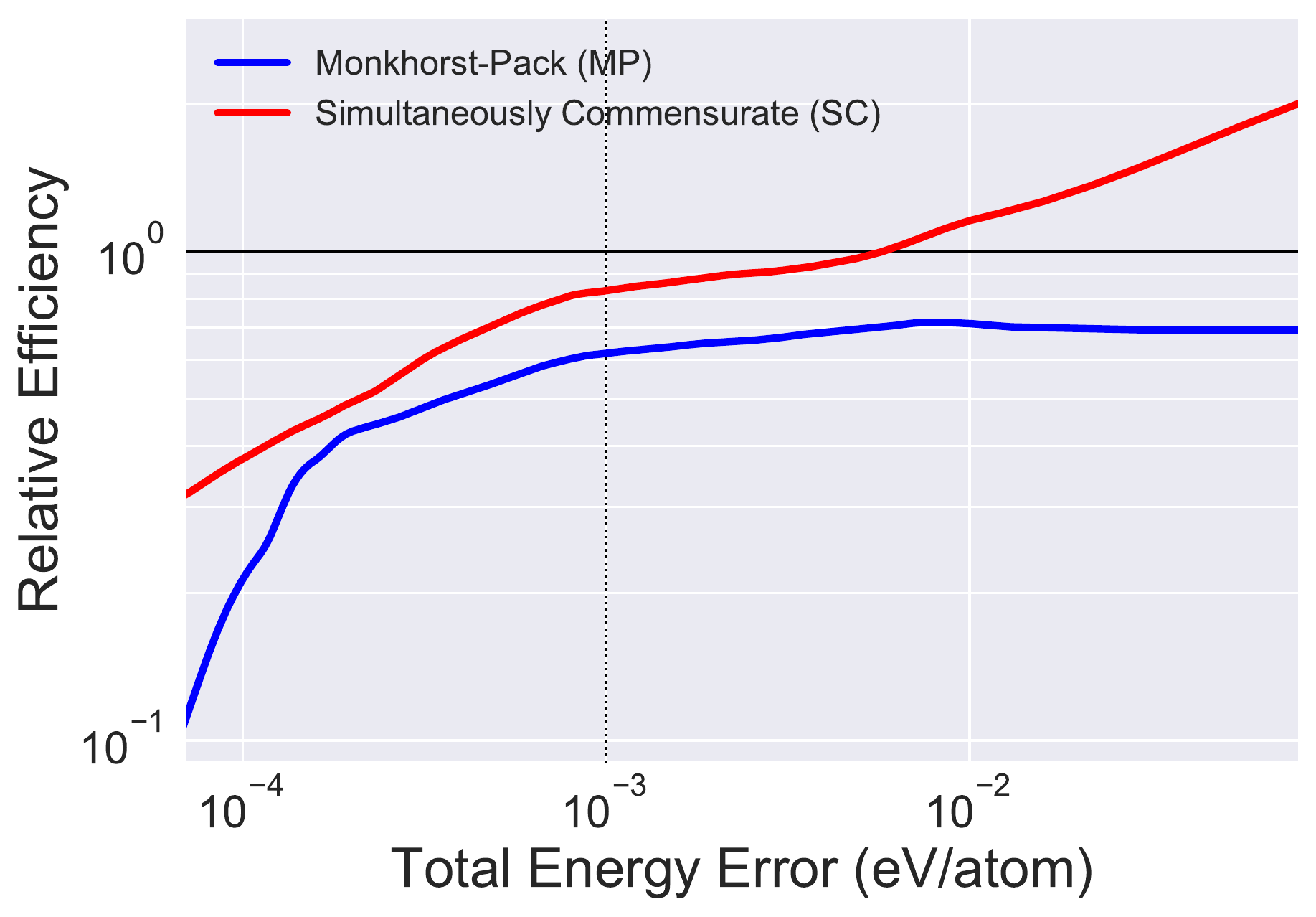}
  \caption{Relative Grid Efficiency. Along the $y$-axis are the ratios
      of the MP and SC efficiencies compared to
      the GR grid efficiency (black horizontal line at $10^0$).
    Total energy error (per atom) is plotted along the $x$-axis and
    decreases to the left. 
    MP and SC grids are generally less efficient than GR grids: at a
    target accuracy of 1 meV/atom, MP grids are 60\%,  and
    SC grids are 20\%, less efficient.}
  \label{fig:relative_performance}
\end{figure}

The efficiency of a \kb-point grid is proportional to the irreducible
\kb-point density required to reach a given accuracy. Comparisons of
efficiencies were made by taking the ratio of the GR trend line to the
SC and MP trend lines of Fig.~\ref{fig:relative_performance}. At
accuracies higher than 5 meV/atom, GR grids are more efficient
(averaged over many structures) than MP and SC grids. As an example,
at a target accuracy of 1 meV/atom, the GR grids are 20\% more
efficient than SC grids and 60\% more efficient than MP grids.

It should be noted that in both Figs.~\ref{fig:all_metal_comb} and
\ref{fig:relative_performance} that MP grids appear to perform worse
at higher densities than at lower densities. Our statistical analysis has
indicated that this behavior is not statistically significant and
likely results from data scarcity for MP grids at high densities. We
believe that with sufficient data for MP grids at these densities the
trend line would continue to run roughly parallel to the GR line
across all densities. However, due to the computational expense of
generating MP grids at such densities we have been unable to
demonstrate this.

\section{Discussion}

The erratic convergence of total energy for metals is attributed
principally to the Fermi surface. Integrating over the occupied
portions of the band structure is equivalent to integrating a
discontinuous band structure over the Brillioun zone; the rapid,
monotonic convergence observed for insulators and semiconductors is
lost because of the surface of discontinuties.

It is perhaps surprising how much the error varies at a given
\kb-point density. The implication is that, when generating databases
of total energies, relatively high \kb-point densities will be
required for accurate comparisons. For example, in
Fig. \ref{fig:all_metal_comb}, \kb-point densities as low as 10s of
\kb-points/\AA$^{-3}$ achieve $10^{-3}$ eV/atom error for some
structures, but to be certain that \emph{all} structures are converged
to the same accuracy densities as high as 5,000 \kb-points/\AA$^{-3}$
are necessary. Given the spread in the data we recommend that a target
density of 5,000 \kb-points/\AA$^{-3}$ be used to reliably achieve
accuracies of $10^{-3}$ eV/atom for metals. However, should another
accuracy be desired, one can simply follow the top edge of the
distribution of points in Fig.~\ref{fig:mueller_comb} to the desired
accuracy and read off the corresponding density.

For reference: a \kb-point density of 5,000 \kb-points/\AA$^{-3}$
corresponds to a linear \kb-point density of 0.058
\AA$^{-1}$ (common input scheme for CASTEP or newer
versions of VASP, \verb=KSPACING= in the \verb=INCAR file=). This is
equivalent to the following Monkhorst-Pack grids or ``\kb-point per
reciprocal atom''(KPPRA) settings for a few pure elements:
\begin{center}
\begin{tabular}{| c | c | c | c | c |}
  \hline
  element & cell divisions & KPPRA \\
  \hline
  W & $43\times43\times43$ & 80,000\\
  \hline
  Cu & $48\times48\times48$ & 110,500 \\
  \hline
  Al & $43\times43\times43$ &  80,000 \\
  \hline
  K & $26\times26\times26$ & 17,500 \\
  \hline
  Ti (2 atoms, hcp) & $41\times41\times21$ & 18,000 \\
  \hline
\end{tabular}
\end{center}
Likely these high numbers will be surprising to most DFT
practitioners---indeed, the current authors found them so---but this
is the message of Fig.~\ref{fig:all_metal_comb} if one wants to be
fully converged in all cases, and not just on average. The large
scatter in the errors for a given density imposes this large penalty
on the practitioner who wishes to have fully converged calculations.

\begin{figure}[t]
  \centering
  \includegraphics[width=8cm]{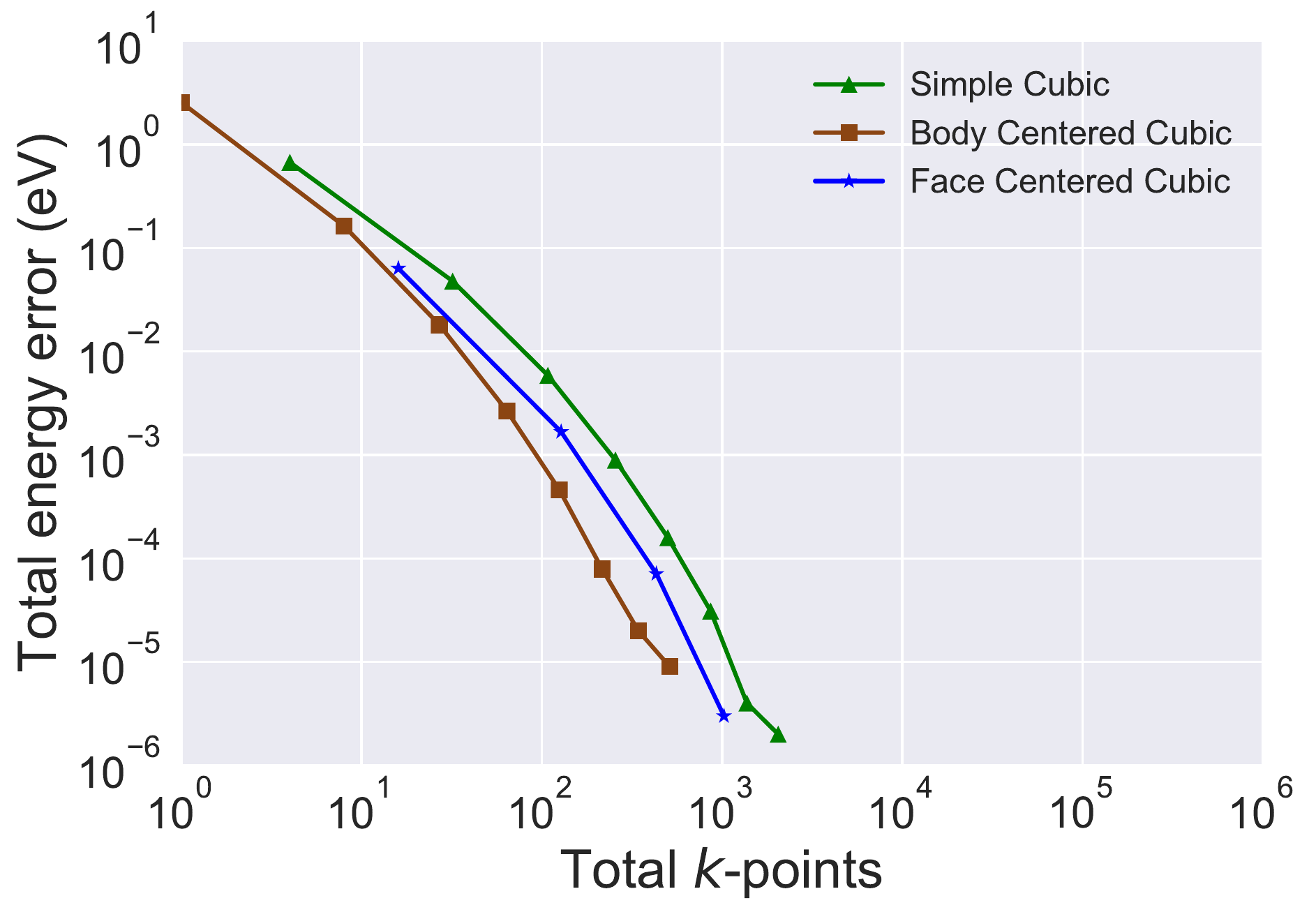}
  \caption{Convergence for silicon by Bravais lattice type of the \kb-point
    grid. The energy convergence remains smooth for GR grids as
    long as the grid is of a single Bravais lattice
    type. Otherwise, some spread in the energy convergence, similar to    
    that observed for metals, is introduced.}
  \label{fig:si_convergence}
\end{figure}

\begin{figure}[t]
  \centering
  \includegraphics[width=8cm]{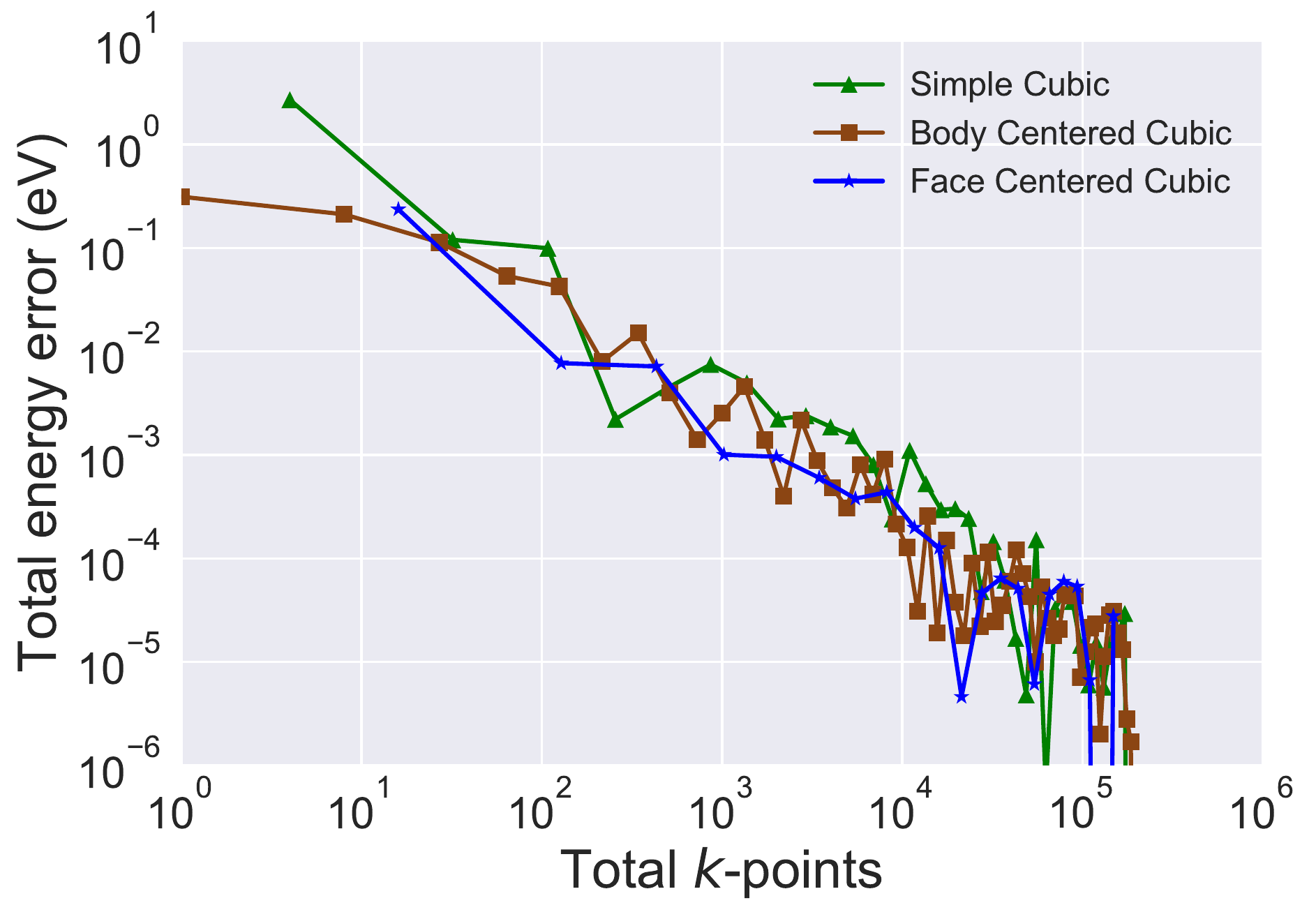}
  \caption{Convergence of aluminum by Bravais lattice type of the \kb-point
    grid. Jaggedness and spread in the energy convergence
    remains for GR grids even after separating the grids by Bravais
    lattice type.}
  \label{fig:al_convergence}
\end{figure}

In our tests of GR grids, we also observed large spread in the energy
convergence of insulators, rather than the typical monotonic
convergence observed for MP grids. This happens because GR grids are
not restricted to a single Bravais lattice type.  Grids of different
lattice types will have different packing fractions and thus converge
at different rates. Figure~\ref{fig:si_convergence} shows the energy
convergence rate of primitive silicon for three Bravais lattice types;
the convergence is monotonic for each type. As expected, body-centered
cubic grids have the fastest convergence. This is because bcc lattices
have the highest packing fraction when Fourier transformed (becoming
fcc). If grids of multiple Bravais lattice types are used, as happens
for GR grids obtained by querying WMM's \kb-point server, spread in
the energy convergence is introduced. To demonstrate that erratic
convergence for metals is not merely due to mixing grids of multiple
Bravais lattice types, we include
Fig.~\ref{fig:al_convergence}. The figure also
  demonstrates that the grid type, i.e., bcc, fcc, or sc, has no
  effect on the convergence.

\section{Conclusion}
GR grids are not intrinsically better than SC or MP grids---that is,
they do not converge more rapidly as a function of \kb-point
density. They are more efficient because they typically have better
symmetry reduction than MP or SC grids, reducing the computational
effort required for GR grids. Also, with GR grids one may
increase the \kb-point density in smaller increments because the set
of possible grids (and thus \kb-point densities) is larger than the
sets of possible grids for SC and MP.

Our tests over more than 7000 structures of varying cell sizes,
shapes, and \kb-point densities demonstrate how erratic \kb-point
convergence is for metals, and how wide the variance can be at a given
\kb-point density, and how this variance grows with increasing
\kb-point densities. These facts should be considered when generating
computational materials databases since greater errors may result from
not using enough \kb-points for a target accuracy. Using GR grids for
non-metals may result in unexpected scatter; when smooth convergence
is desired, we advise that GR grids of a single Bravais lattice type
be utilized.

\section{Acknowledgments}

The authors are grateful to Shane Reese who helped with the loess
regression and statistical analysis of the data shown
Fig.~\ref{fig:all_metal_comb}. The authors are grateful to Tim Mueller
and Georg Kresse for helpful discussions. This work was supported
under: ONR (MURI N00014-13-1-0635).

The raw and processed data required to reproduce these findings are
available to download from
\url{https://github.com/wsmorgan/GR\_Grid\_Comparisons}%{https://github.com/wsmorgan/GR\_Grid\_Comparisons}.

\section{Appendix}

\subsection{Simultaneously Commensurate Grid Construction}

\begin{figure}
  \centering
  \includegraphics[width=8cm]{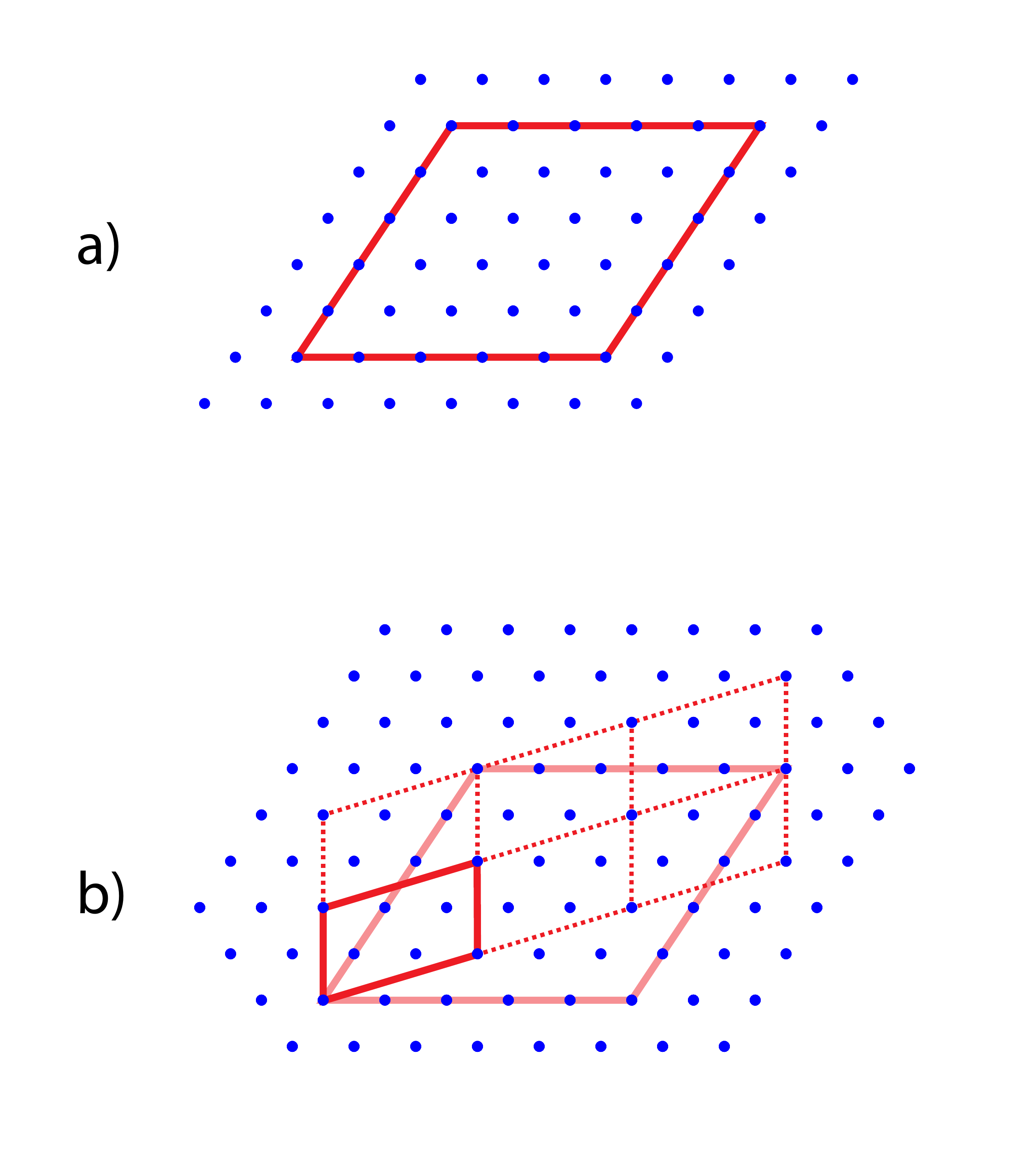}
  \caption{An example of a SC grid. In part a of the figure a grid is
    created that is commensurate with the parent cell. In part b the
    supercell of the parent is added, as can be seen in part b the
    grid is commensurate with both the parent and the supercell.}
  \label{fig:commensurate-plot}
\end{figure}

\begin{figure}
  \centering
  \includegraphics[width=8cm]{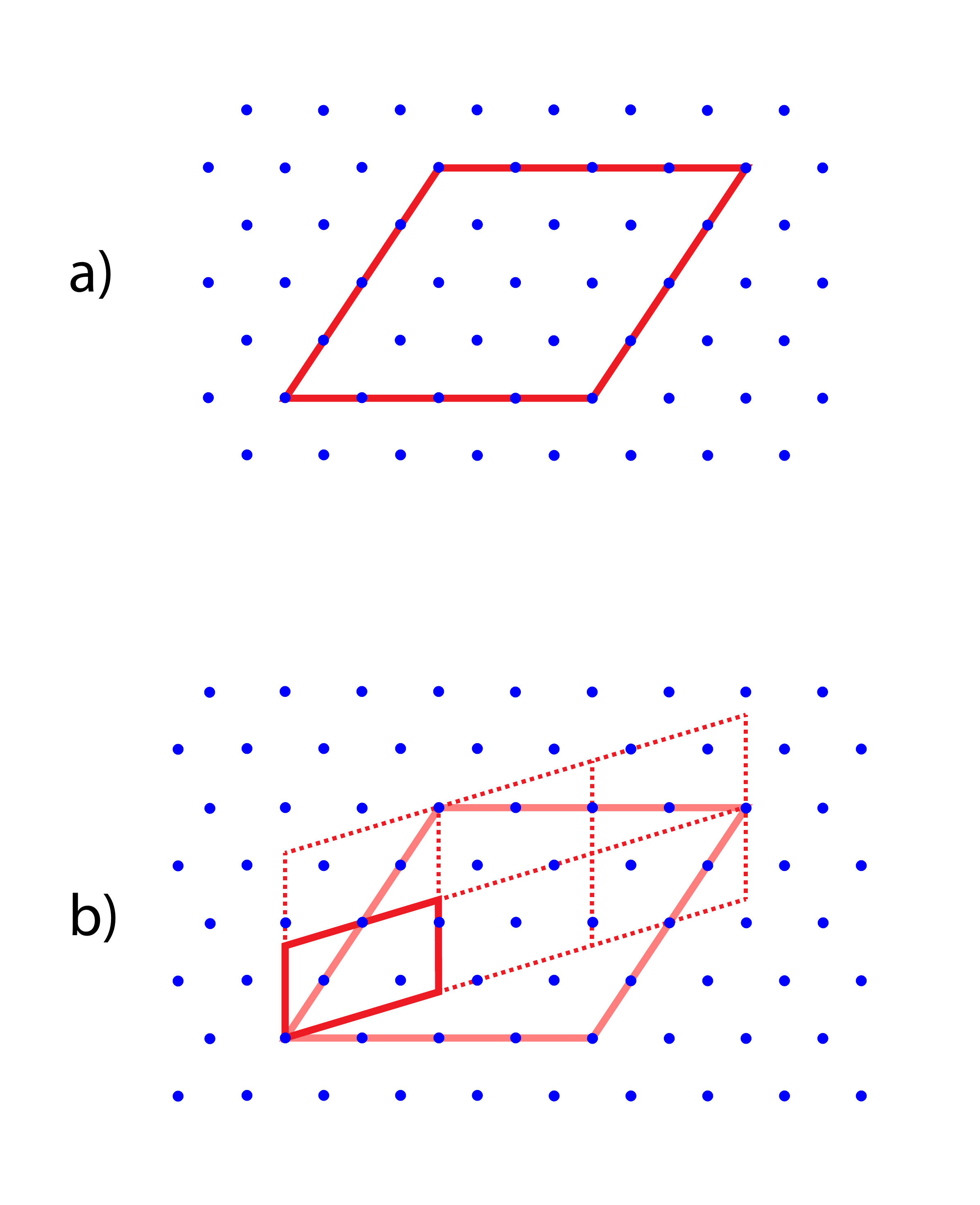}
  \caption{An example of a SC grid that is not commensurate. In part a
    of the figure a grid is created that is commensurate with the
    paernt cell. In part b the supercell of the parent is added, as
    can be seen in part b the grid is not periodic, i.e.,
    commensurate, with the supercell. This grid would therefore be
    invalid for use to integrate the supercell.}
  \label{fig:commensurate-plot2}
\end{figure}

A \emph{simultaneously commensurate} (SC) grid is useful for
calculating formation enthalpies when the target structure is a
derivative superstructure of a parent structure. (Obviously this is a
convenient method when computing enthalpies for cluster expansion
studies because the training structures are superstructures of the
parent.)  When SC grids are used, the absolute convergence with
respect to \kb-density is not faster than for other grids but the
\emph{relative convergence} can be faster because of error
cancellation---both the parent structure and the derivative
superstructure have exactly the same grid. The idea is illustrated in
Fig~\ref{fig:commensurate-plot}. In panel a) we divide up the
reciprocal unit cell of the parent lattice (red parallelogram) into a
uniform grid of \kb-points (blue points). We then place the same grid from
the parent cell on the supercell, as in panel b). If we have chosen a
SC grid, it is clearly
periodic for the supercell as well as the parent. Only those grids
that are commensurate with both the parent cell and supercell can be
used to integrate both cells. Fig. ~\ref{fig:commensurate-plot2}
shows an example of an incommensurate grid. When the grid of the parent
cell is place over the reciprocal cell of the supercell, the grid is
\emph{not periodic}---translations of the supercell (dotted lines) are sampled
differently by the grid.

For our crystals that have cubic parent cells, an initial set of
commensurate bcc, fcc, and sc grids were generated. A subset of those
grids that were commensurate with each supercell were used to do
calculations of the various crystal structures. For hexagonal
crystals, a similar procedure was followed except only hexagonal grids
were used.

%% \bibliography{bib.bib}{} %\bibliographystyle{unsrt}


\clearpage
\onecolumngrid
\section{Supplemental Materials}

\begin{figure}[h]
  \centering
  \includegraphics[width=18cm]{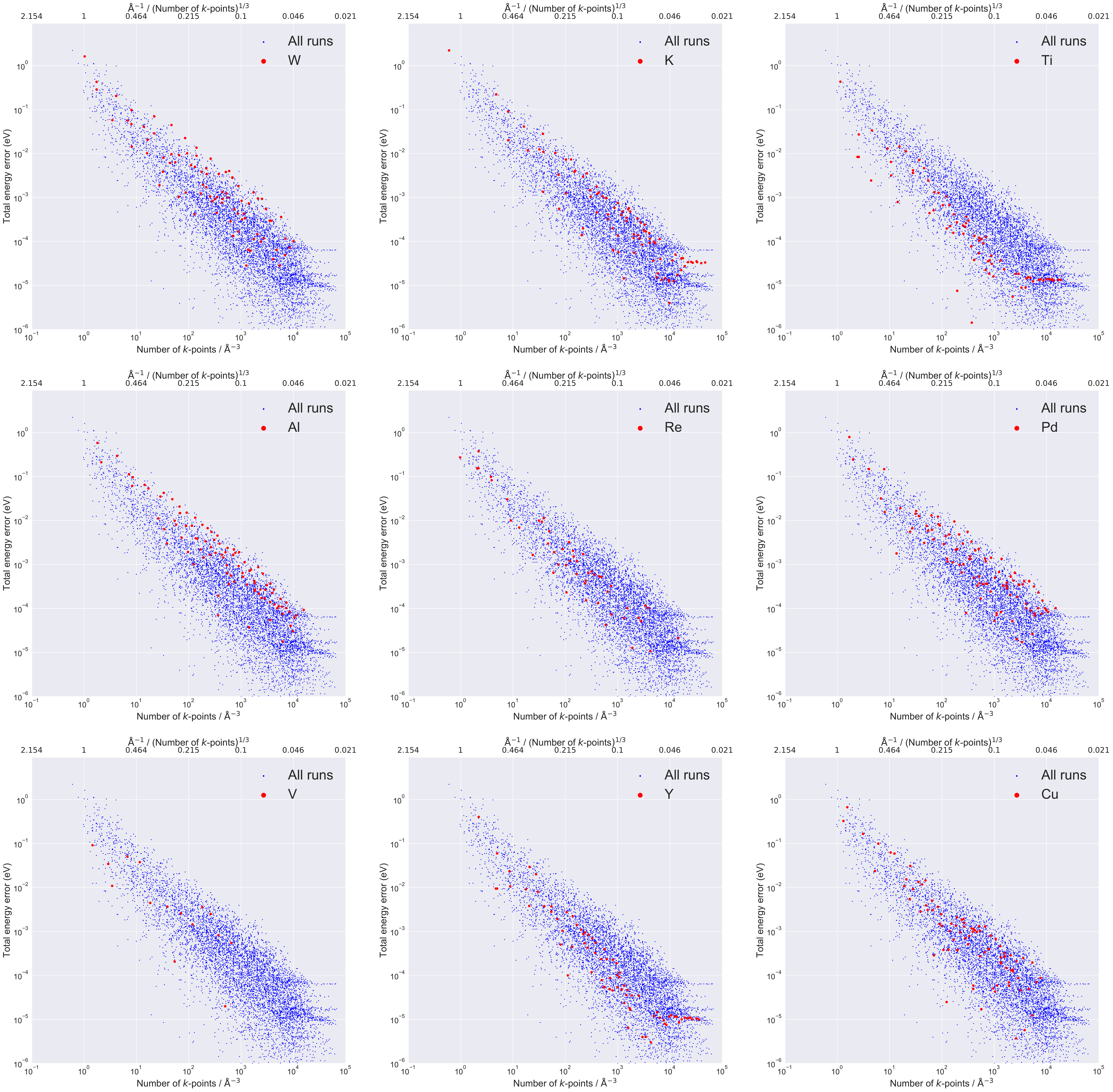}
  \caption{Plots of the convergence of each elements primitive cell
    overlaid on the convergence of all sysetms for comparison.}
\end{figure}
\end{document}